\documentstyle[iopconf1,epsf]{article}
\begin{document}
\title{
{\small 
\begin{flushright}
CERN-TH/99-249\\
THES-TP/99-10\\
hep-ph/9908373\\
August 1999
\end{flushright}}
CP VIOLATION IN THE HIGGS SECTOR OF THE MSSM}\footnotetext[1]{To appear in
the proceedings of ``Beyond the Desert,'' ed.\ H.V.
Klapdor-Kleingrothaus, Castle Ringberg, Tegernsee, 6--12 June 1999,
Germany.}

\author{Apostolos Pilaftsis}

\affil{Theory Division, CERN, CH-1211 Geneva 23, Switzerland\\[0.1cm]
                       and\\[0.1cm]
Department of Theoretical Physics, University of Thessaloniki,\\
GR 54006 Thessaloniki, Greece}

\beginabstract  Recently, it has  been   found that the tree-level  CP
invariance of the Higgs potential  in the MSSM  can be sizeably broken
by  loop  effects  due   to  soft-CP-violating  trilinear interactions
involving third  generation   scalar quarks.   These soft-CP-violating
couplings may be constrained by considering new two-loop contributions
to the  electron and neutron  EDMs.  The phenomenological consequences
of such a minimal supersymmetric  scenario of explicit CP violation at
present and future colliders are briefly discussed.  \endabstract

\section{Introduction}

Many studies  have  been  devoted  to  understand  the  origin of  the
observed CP asymmetry in the kaon system.  In the existing literature,
two generically different scenarios are known to describe CP violation
in the Higgs sector of a quantum field theory.  In the first scenario,
CP   invariance  is broken explicitly   by  complex bilinear  terms or
quartic  couplings  that involve Higgs  doublets  in an extended Higgs
sector.  Such a scenario  predicts a  CP-violating scalar-pseudoscalar
mixing already at the tree level.   Another interesting scenario is to
have the CP symmetry of the  Lagrangian be spontaneously broken by the
ground   state  of  the Higgs   potential, while    all parameters and
couplings are real and respect CP invariance.  To realize one of these
two schemes, one  needs  to extend the Higgs  sector  of the  Standard
Model (SM) at least by one additional Higgs doublet.  The most minimal
supersymmetric   extension   of the   SM, the   so-called   MSSM, with
$R$-parity invariance, cannot realize any  of the above two schemes at
the  tree level, despite  the fact that  the model  contains two Higgs
doublets. Beyond the Born approximation,  the MSSM Higgs potential can
break  the CP symmetry  either  spontaneously \cite{NM}  or explicitly
\cite{AP}.  The  spontaneous CP-violating MSSM  predicts  a very light
Higgs scalar below   10    GeV, which  is ruled  out    experimentally
\cite{Alex}.

Recently, it has been found \cite{AP,PW}, however, that the tree-level
CP invariance of  the MSSM Higgs potential  can be maximally broken at
the one-loop  level if soft-CP-violating Yukawa interactions involving
stop and sbottom  quarks are present in  the theory.   As an immediate
consequence, the small tree-level mass difference between the heaviest
Higgs boson and the CP-odd scalar may be lifted considerably through a
large CP-violating  scalar-pseudoscalar mass  term \cite{AP,PW,Demir}. 
This radiative scenario of   explicit CP violation constitutes  a very
interesting possibility within the framework of the MSSM, and we shall
briefly discuss its main  phenomenological consequences at present and
planned collider machines.

\section{The effective CP-violating Higgs potential}

It is known that the MSSM  introduces several new CP-odd phases in the
theory   that   are  absent   in   the   Standard  Model   \cite{EDM}.
Specifically,  the  new CP-odd  phases  may  come  from the  following
parameters: (i) the parameter $\mu$ that describes the bilinear mixing
of     the    two     Higgs    chiral     superfields;     (ii)    the
soft-supersymmetry-breaking  gaugino masses  $m_\lambda$ for  which we
assume to have a common phase at the unification point; (iii) the soft
bilinear  Higgs-mixing mass  $m^2_{12}$; and  (iv) the  soft trilinear
Yukawa couplings $A_f$  of the Higgs particles to  the scalar partners
of matter  fermions.  Not  all phases of  the four  complex parameters
mentioned  above are  physical, i.e.\  two  phases may  be removed  by
suitable redefinitions  of the fields.   For example, one  can rephase
one of the  Higgs doublets and the gaugino fields  $\lambda$, in a way
such that arg($\mu$) and arg($A_f$) are the only physical CP-violating
phases in the MSSM.

An immediate consequence of  Higgs-sector CP violation  in the MSSM is
the presence of mixing-mass terms between the CP-even and CP-odd Higgs
fields \cite{AP}.  Thus,  one finds  a $(4\times 4)$-dimensional  mass
matrix for the neutral  Higgs  bosons.  In the  weak basis  $(G^0,  a,
\phi_1, \phi_2)$, where  $G^0$ is  the  Goldstone field,  the  neutral
Higgs-boson mass matrix ${\cal M}^2_0$ takes on the form
\begin{equation}
  \label{NHiggs}
{\cal M}^2_0 \ =\ 
\left(\begin{array}{cc} \widehat{\cal M}^2_P  & {\cal M}^2_{PS} \\
                {\cal M}^2_{SP} &  {\cal M}^2_S \end{array} \right)\, ,
\end{equation}
where   ${\cal  M}^2_S$  and   $\widehat{\cal  M}^2_P$   describe  the
CP-conserving transitions  between scalar and  pseudoscalar particles,
respectively,  whereas  ${\cal   M}^2_{PS}  =  ({\cal  M}^{2}_{SP})^T$
describes    CP-violating   scalar-pseudoscalar    transitions.    The
characteristic size  of these  CP-violating off-diagonal terms  in the
Higgs-boson mass matrix was found to be \cite{AP,PW}
\begin{eqnarray}
M^2_{SP} & \simeq & {\cal O} \left( \frac{ m_t^4}{v^2} 
\frac{|\mu| |A_t|}{32 \pi^2M_{\rm SUSY}^2} \right) \sin \phi_{\rm CP} 
\nonumber\\
&&\times\, \left(6,\ \frac{|A_t|^2}{M_{\rm SUSY}^2}\, ,\ \frac{|\mu|^2}
{\tan\beta\, M_{\rm SUSY}^2}\,,\ \frac{\sin 2\phi_{\rm CP}}{\sin
  \phi_{\rm CP}}\, \frac{|\mu||A_t|}{M_{\rm SUSY}^2} \right),
\end{eqnarray}  
where the last bracket summarizes  the relative sizes of the different
contributions, and  $\phi_{\rm CP} = {\rm   arg}(A_t \mu)\, +\,  \xi$. 
The parameter $\xi$ is the relative phase between the two Higgs vacuum
expectation values which is induced radiatively in the $\overline{MS}$
scheme \cite{AP,PW}. 

It worth  commenting  on the renormalization-scheme dependence  of the
phase  $\xi$.  For  example, one may   adopt a renormalization scheme,
slightly different from the $\overline{MS}$ one, in which $\xi$ is set
to zero order by order in perturbation theory  \cite{AP}.  This can be
achieved by requiring for the bilinear Higgs-mixing mass $m^2_{12}$ to
be real  at  the tree  level, but  receive an  imaginary counter-term,
${\rm Im}  m^2_{12}$,  at higher  orders,  which is determined by  the
vanishing of  the CP-odd tadpole    parameter of the  would-be  CP-odd
scalar for $\xi = 0$. This is a peculiarity of the CP-odd CT ${\rm Im}
(m^2_{12} e^{i\xi})$, which appears as an independent parameter in the
effective action,  i.e.\  its  scheme   of  renormalization  does  not
directly affect the renormalization scheme of other physical kinematic
parameters of  the theory to one-loop,  such as Higgs-boson masses and
$\tan\beta$.  In fact, as was  shown in \cite{AP}, physical transition
amplitudes between different Higgs  states such as scalar-pseudoscalar
transitions, are independent of the renormalization subtraction point.
The   above   scheme of  renormalization     is  rather useful,  since
unnecessary  $\xi$-dependent  phases in  the   tree-level chargino and
neutralino mass matrices may be avoided.

\begin{figure}
   \leavevmode
 \begin{center}
   \epsfxsize=12.0cm
    \epsffile[0 0 539 652]{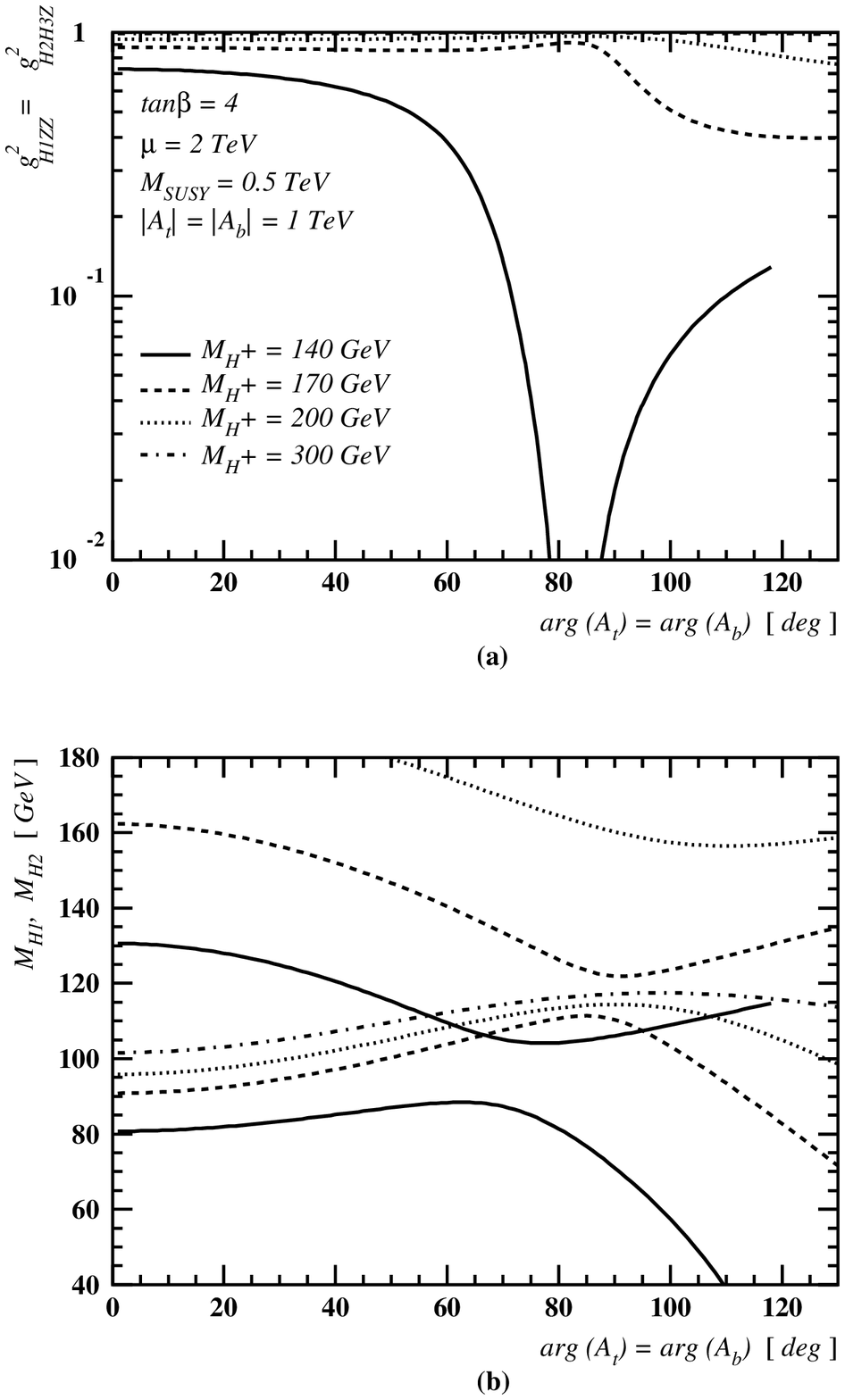}
 \end{center}
 \vspace{-0.7cm} 
\caption{Numerical estimates of (a) $g^2_{H_1ZZ}$ and (b) $M_{H_1}\le
  M_{H_2}$ as a function of ${\rm arg} (A_t)$.}
\label{fig:scp7}
\end{figure}
The CP-violating effects can become substantial if $|\mu|$ and $|A_t|$
are larger  than the  average of the  stop  masses, denoted as $M_{\rm
  SUSY}$.  For  example,  the  off-diagonal   terms of  the    neutral
Higgs-mass matrix  may be of order   $(100\ {\rm GeV})^2$,  for $|\mu|
\simeq |A_t| \stackrel{<}{{}_\sim} 3 M_{\rm SUSY}$, and $\phi_{\rm CP}
\simeq 90^\circ$.  These   potentially  large mixing effects lead   to
drastic  modifications of the predictions  for the neutral Higgs-boson
masses and  for the couplings of the  Higgs states to the gauge bosons
\cite{PW}.  As can be seen from Fig.\ \ref{fig:scp7}, the effect of CP
nonconservation  on   the lightest  Higgs  boson  and   on its related
couplings to the  gauge  bosons is  only important for  relatively low
values   of $M_{H^+}$, e.g.\  $M_{H^+}  \stackrel{<}{{}_\sim} 170$ GeV
\cite{PW}. The upper  limit on the  lightest Higgs-boson mass does not
change, as the relevant stop mixing  parameter entering the definition
of $M_{H_1}$ is now given by
\begin{equation}
  \label{tildeat}
|\tilde{A}_t|\ =\ |
A_t - \mu^* /\tan\beta|\, .
\end{equation}
Notice   that the scenario   we have used   in Fig.\ \ref{fig:scp7} is
compatible with experimental upper  limits on the electron and neutron
electric dipole  moments   (EDMs).   These  EDM  constraints will   be
discussed in the next section.

\section{Two-loop EDM constraint}

\begin{figure}
   \leavevmode
 \begin{center}
   \epsfxsize=10.0cm
   \epsfysize=6.0cm
    \epsffile[100 500 500 700]{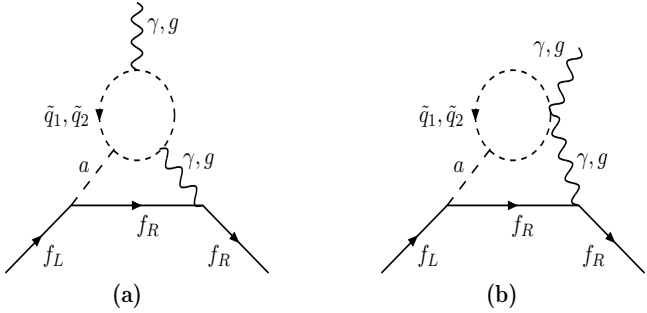}
 \end{center}
 \vspace{-2.cm} 
\caption{Two-loop contribution to EDM and CEDM in supersymmetric
  theories (mirror graphs are not displayed.)}\label{f1}
\end{figure}

The MSSM generally   gives  large contributions to  the  electron  and
neutron EDMs,   coming  from  squarks  of   the first   two   families
\cite{EDM,CKP}. Even if the first two families of squarks are arranged
so  as to  give small FCNC   and EDM  effects \cite{GD},  the two-loop
graphs \cite{CKP} shown in Fig.\ \ref{f1} may even dominate by several
orders  of  magnitude  over all   other  one-,  two-   and  three-loop
contributions, thereby   significantly  constraining   Higgs-sector CP
violation.  In the SUSY scenario, with a large CP-violating phase only
in the  third  family $A_\tau  = A_t   =  A_b =  A$,  the CP-violating
Lagrangian,
\begin{equation}
  \label{Lcp}
{\cal L}_{\rm CP}\ =\ -\, \xi_{\tilde{f}}
v\, a\, (\tilde{f}_1^* \tilde{f}_1\, -\,
\tilde{f}_2^* \tilde{f}_2)\
+\ \frac{ig_w m_f}{2M_W}\, R_f\, a\,\bar{f}\gamma_5 f\, ,
\end{equation}
gives rise to a  new EDM contribution to  the neutron and electron. In
Eq.\ (\ref{Lcp}),  $a$  is the  would-be CP-odd  Higgs  boson, $M_W  =
{1\over2}g_w v$ is  the $W$-boson mass,  $\tilde{f}_{1,2}$ are the two
mass-eigenstates   of  the  third-family    squarks,  $R_b=\tan\beta$,
$R_t=\cot\beta$,  and    $\xi_f$ is  a   model-dependent  CP-violating
parameter. In the MSSM, only $\tilde{t}$  and $\tilde{b}$ are expected
to give  the  largest contributions,   as the quantities  $\xi_f$  are
Yukawa-coupling enhanced, viz.
\begin{equation}
  \label{xiq}
\xi_{\tilde{f}}\ =\ -\, R_f\, \frac{\sin 2\theta_f m_f {\rm Im} ( \mu
  e^{i\delta_f})}{\sin\beta\, \cos\beta\, v^2}\,,
\end{equation}
where   $\delta_f =  {\rm  arg}  (A_f  - R_f  \mu^*)$, and $\theta_t$,
$\theta_b$ are the mixing angles between weak  and mass eigenstates of
$\tilde{t}$ and $\tilde{b}$, respectively. Further details of the 
calculation may be found in \cite{CKP}.

\begin{figure}
   \leavevmode
  \begin{center}
  \epsfxsize=12.0cm
    \epsffile[0 0 539 652]{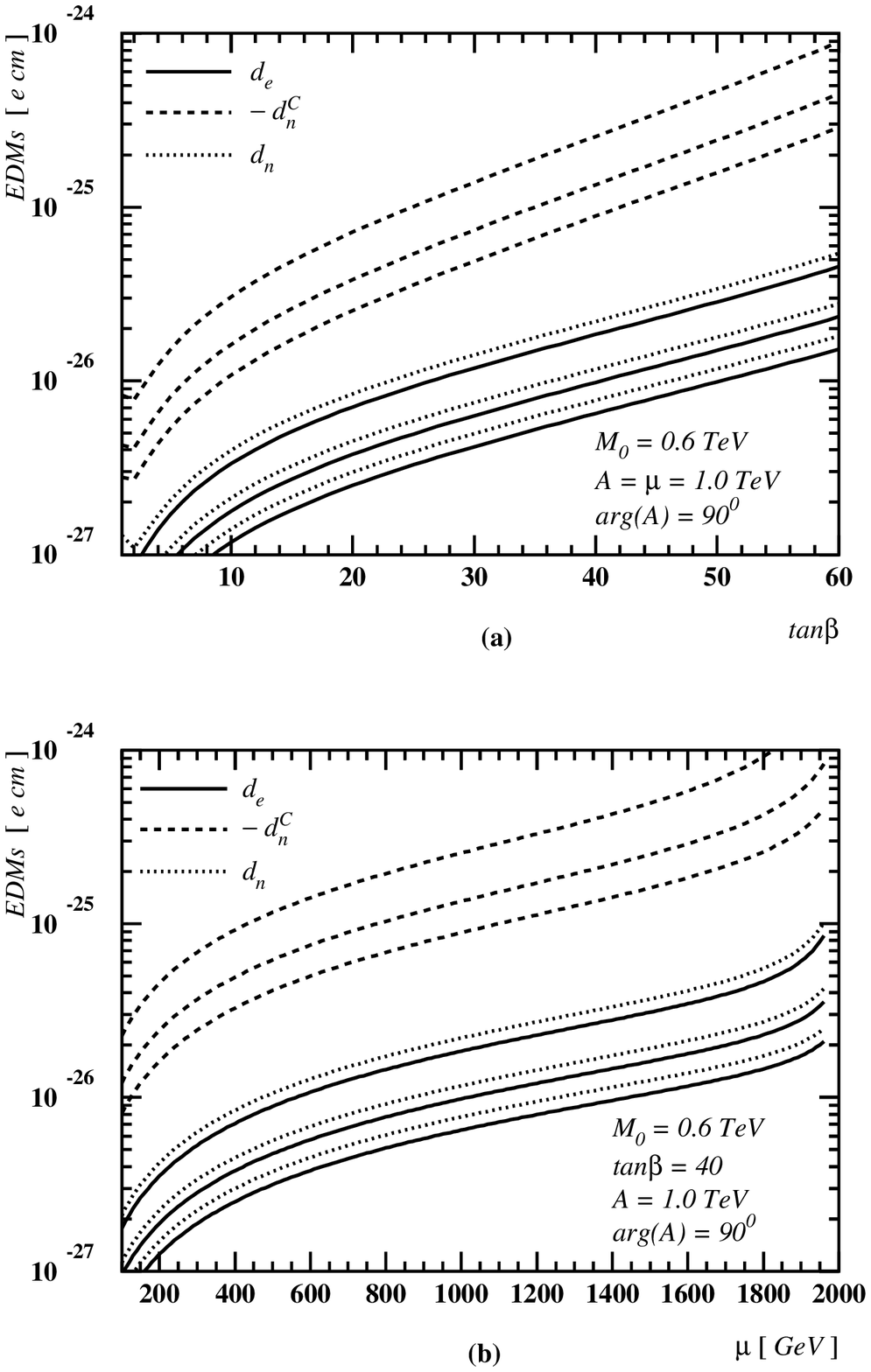}
 \end{center}
 \vspace{-0.7cm}
\caption{Numerical estimates of EDMs. Lines of the same type
from the upper to the lower one correspond to $M_a = 100,\ 300,\
500$ GeV, respectively.}\label{f2}
\end{figure}

Fig.\ 2 shows the  dependence of the  EDMs $d_e$ (solid line), $d^C_n$
(dashed  line), and $d_n$ (dotted  line) on $\tan\beta$ and $\mu$, for
three different masses of the would-be  CP-odd Higgs boson $a$, $M_a =
100,  300, 500$ GeV.  Since   the coupling of    the $a$ boson to  the
down-family fermions   such as  the electron  and  $d$   quark depends
significantly on $\tan\beta$, we  find a substantial increase of $d_n$
and $d_e$ in the large $\tan\beta$ domain (see Fig.\ 2(a)).  As can be
seen   from  Fig.\  2(b),  EDMs also   depend   on $\mu$   through the
$a\tilde{f}^*\tilde{f}$ coupling  in Eq.\  (\ref{Lcp}).  Note that the
numerical predictions for the  size of EDMs do  not depend on the sign
of $\mu$ for ${\rm arg}(A) = 90^0$.  {}From our numerical analysis, we
can exclude large $\tan\beta$  scenarios, {\em i.e.}, $40< \tan\beta <
60$ with $\mu,\ A > 0.5$ TeV, $M_a \le 0.5$  TeV, and large CP phases. 
Nevertheless,   the   situation   is different   for   low $\tan\beta$
scenarios,  {\em e.g.}\   $\tan\beta \stackrel{\displaystyle  <}{\sim}
20$, where the two-loop Barr-Zee-type contribution to EDMs is not very
restrictive for  natural values of parameters   in the MSSM.  Finally,
EDMs display a mild linear dependence on the mass of the $a$ boson for
the range of  physical interest,  $0.1 < M_a   \stackrel{\displaystyle
  <}{\sim} 1$ TeV.

\section{Higgs phenomenology and CP violation}

The main  effect of Higgs-sector  CP violation  is the modification of
the couplings  of  the Higgs  bosons to fermions  and the  $W$ and $Z$
bosons,  i.e.\ $ffH_i$, $WWH_i$,  $ZZH_i$  and $ZH_iH_j$. The modified
effective Lagrangians are given by
\begin{eqnarray}
  \label{Hiff}
{\cal L}_{H\bar{f}f} &=& -\, \sum_{i=1}^3 H_{(4-i)}\,
\Big[\, \frac{g_w m_{d}}{2M_W c_\beta}\, \bar{d}\,( O_{2i}\, -\,
is_\beta O_{1i}\gamma_5 )\, d\nonumber\\
&& +\, \frac{g_w m_u}{2M_W s_\beta}\, \bar{u}\,( O_{3i}\, -\,
ic_\beta O_{1i}\gamma_5 )\, u\, \Big]\, ,\\
  \label{HVV}
{\cal L}_{HVV} &=&  g_w M_W\, ( c_\beta O_{2i}\, +\, s_\beta
O_{3i} )\, \Big(\, H_{(4-i)} W^+_\mu W^{-,\mu}\nonumber\\
&&+\ \frac{1}{2c^2_w}\, H_{(4-i)} Z_\mu Z^\mu\, \Big)\, ,\\
  \label{HHZ}
{\cal L}_{HHZ} &=& \frac{g_w}{4c_w}\, 
\Big[\, O_{1i}\, ( c_\beta O_{3j}\, -\, s_\beta
O_{2j} )\, -\, O_{1j}\, ( c_\beta O_{3i}\, -\, s_\beta
O_{2i} )\, \Big]\, \nonumber\\
&&\times\, Z^\mu\, ( H_{(4-i)}\, \!\!
\stackrel{\leftrightarrow}{\vspace{2pt}\partial}_{\!\mu} H_{(4-j)} )\, ,
\end{eqnarray}
where   $c_w =   M_W/M_Z$ and $\stackrel{\leftrightarrow}{\vspace{2pt}
  \partial}_{\!   \mu}\    \equiv\ \stackrel{\rightarrow}{\vspace{2pt}
  \partial}_{\!      \mu}    -      \stackrel{\leftarrow}{\vspace{2pt}
  \partial}_{\! \mu}$.  Note that the coupling of the $Z$ boson to two
real scalar fields is forbidden due to Bose symmetry.

\begin{figure}
   \leavevmode
 \begin{center}
   \epsfxsize=12.0cm
    \epsffile[0 0 539 652]{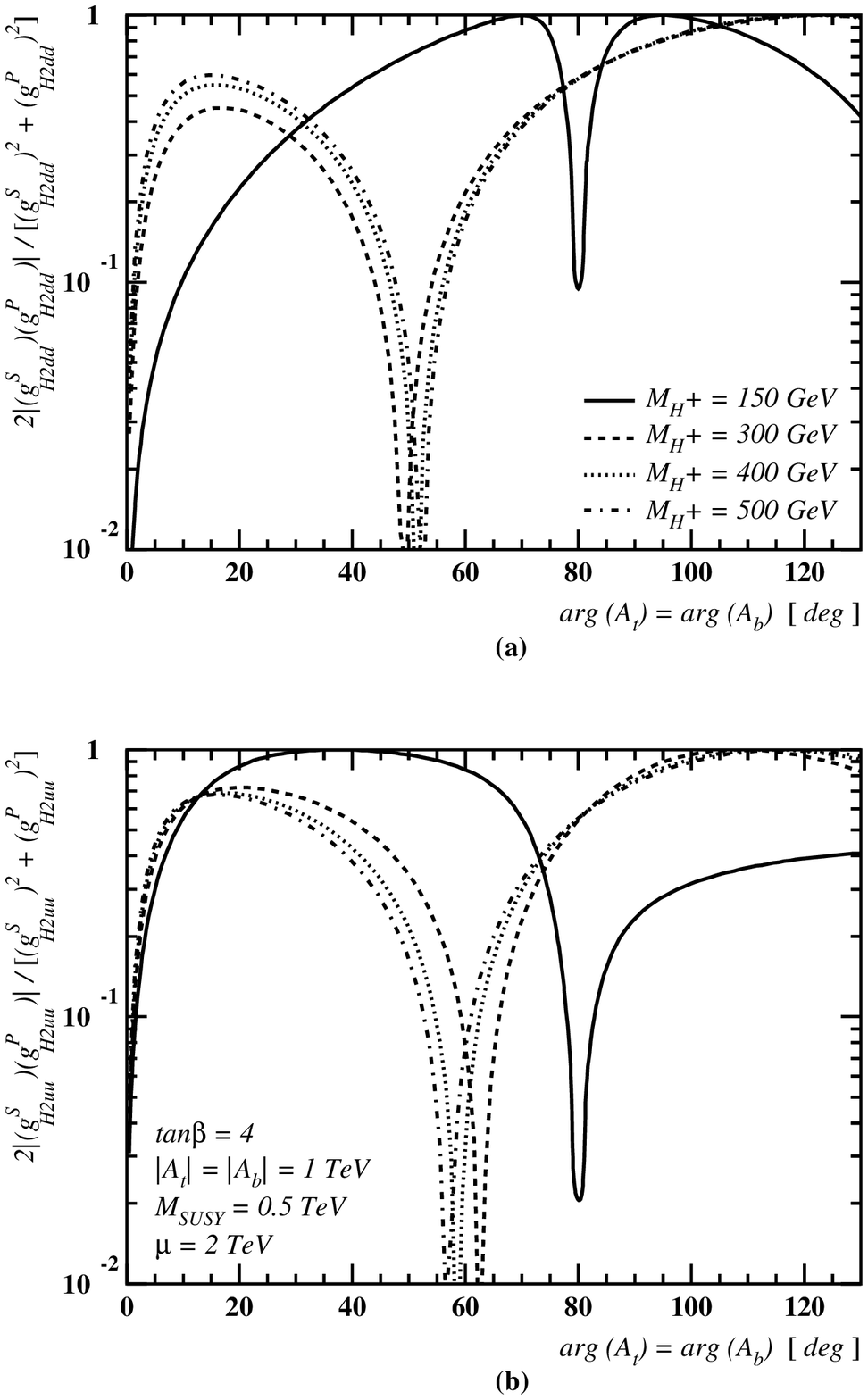}
 \end{center}
 \vspace{-0.5cm} 
\caption{Numerical estimates of (a) $2|(g^S_{H_2dd})\,
(g^P_{H_2dd})|/
  [(g^S_{H_2dd})^2 + (g^P_{H_2dd})^2]$ and (b)
  $2|(g^S_{H_2uu})\, (g^P_{H_2uu})|/[(g^S_{H_2uu})^2 + (g^P_{H_2uu})^2]$
  as a function of ${\rm arg} (A_t )$.}
\label{fig:scp17}
\end{figure}

We shall  now discuss a generic example  in order to better understand
the  effect of Higgs-sector  CP violation on the  mass of the lightest
Higgs boson.   We consider an  intermediate-$\tan\beta$ scenario, with
$\tan\beta =4$, $M_{\rm SUSY}= 0.5$ TeV, $A_t = A_b =  1$ TeV and $\mu
= 2$ TeV.  {}From Fig.\ \ref{fig:scp7} and for $M_{H^+} = 140$ GeV, we
observe  that there exist regions   for which the lightest Higgs-boson
mass $M_{H_1}$ is as  small as 60--70  GeV and the $H_1ZZ$ coupling is
small enough  for the $H_1$  boson to escape   detection at the latest
LEP2 run with $\sqrt{s} = 189$ GeV.  In this scenario, the $H_2$ boson
is too heavy to be detected through the $H_2ZZ$ channel.  Furthermore,
we find that  either the coupling $H_1H_2Z$ is  too small or $H_2$  is
too heavy to allow Higgs detection in the $H_1H_2Z$ channel \cite{PW}.
An upgraded Tevatron machine has  the potential capabilities to  close
most of such experimentally open windows.

It   is worth stressing  that   the CP-violating MSSM  Higgs potential
retains its enhanced  predictive  power in the lightest  Higgs ($H_1$)
sector.  As  was mentioned in Section  2,  CP violation decouples from
the $H_1$   sector for large  values   of $M_{H^+}\approx M_{H_2} \sim
M_{H_3}$, but it is  sizeable for $M_{H^+}  \stackrel{<}{{}_\sim} 170$
GeV \cite{PW}.  However, for  much larger  $H^+$  masses, CP violation
does not decouple in   the $H_2H_3$ system \cite{AP,PW,Demir},  giving
rise to large CP-violating effects  in the $H_{2,3}uu$ and $H_{2,3}dd$
couplings. In fact, the necessary condition  for resonant CP violation
through scalar-pseudoscalar ($HA$) mixing is \cite{APRL}
\begin{equation}
  \label{cond}
2|\Pi_{HA} (s)|\ \stackrel{>}{{}_\sim}\ | M^2_H - M^2_A - \Pi_{HH} (s)
+ \Pi_{AA} (s)|\, ,
\end{equation}
at  $s\approx  M^2_H  \approx  M^2_A$, where  $\Pi_{HH,HA,AA}$  denote
renormalized self-energy  transitions of Higgs scalars  of definite CP
parity, namely $H$ has CP parity  $+1$ and $A$ has CP parity $-1$.  As
was  shown in  \cite{AP},  the condition  (\ref{cond}) is  comfortably
satisfied  within  the framework  of  MSSM. Furthermore,  CP-violating
effects induced by radiative corrections to the Higgs-fermion vertices
may also  be important, particularly  for large values  of $\tan\beta$
\cite{PW}.

Higgs-sector  CP violation  may also  be tested at  muon  colliders by
looking at observables of the kind \cite{APRL}
\begin{eqnarray}
  \label{muacp}
{\cal A}^{\mu}_{\rm CP} &=& \frac{\sigma (\mu^-_L\mu^+_L\to f \bar{f})\, -\, 
\sigma (\mu^-_R \mu^+_R\to f \bar{f}) }{\sigma (\mu^-_L\mu^+_L\to
f \bar{f})\,  +\, \sigma (\mu^-_R\mu^+_R\to f \bar{f}) }\, ,\\
  \label{facp}
{\cal A}^{f}_{\rm CP} &=& \frac{\sigma (\mu^-\mu^+\to f_L \bar{f}_L)\, -\, 
\sigma (\mu^-\mu^+\to f_R \bar{f}_R )}{\sigma (\mu^-\mu^+\to
f_L \bar{f}_L)\,  +\, \sigma (\mu^-\mu^+\to f_R \bar{f}_R)}\, ,
\end{eqnarray}
where $f$ may be top  or bottom quarks. The former observable requires
polarization of the initial muons.  If the facility of polarization is
not available  at muon colliders,  one may still observe  CP violation
through the second observable  and reconstruct the polarization of the
final  fermions by  looking at  the angular  momentum  distribution of
their decay products \cite{GG}.   The magnitudes of these CP-violating
observables  strongly  depend  on the  expressions  $2|(g^S_{H_iff})\,
(g^P_{H_iff})|/[(g^S_{H_iff})^2 + (g^P_{H_iff})^2]$ (see also Fig.\
\ref{fig:scp17}), where
\begin{eqnarray}
  \label{gHff} g^S_{H_1uu} \!\!\!&=&\!\!\! O_{33}/s_\beta\,,\quad
g^P_{H_1uu} = O_{13}\, \cot\beta\,,\quad g^S_{H_2uu} =
O_{32}/s_\beta\,,\nonumber\\ g^P_{H_2uu} \!\!\!&=&\!\!\! O_{12}\,
\cot\beta\,,\quad g^S_{H_1dd} = O_{23}/c_\beta\,,\quad g^P_{H_1dd} =
O_{13}\, \tan\beta\,,\nonumber\\ g^S_{H_2dd} \!\!\!&=&\!\!\!
O_{22}/c_\beta\,,\quad g^P_{H_2dd} = O_{12}\, \tan\beta\, .\qquad
\end{eqnarray}

To summarize,  the MSSM with  radiatively induced CP violation  in the
Higgs  sector  is  a   very  predictive  theoretical  framework,  with
interesting  consequences  on   collider  experiments  \cite{CMW},  CP
asymmetries  in   $B$-meson  decays  \cite{Bmeson},   and  dark-matter
searches \cite{DM}.

I  wish to thank  Darwin Chang,  Wai-Yee  Keung and Carlos  Wagner for
collaboration. I also thank the Theory Groups of SLAC and FERMILAB for
their kind hospitality.

\end{document}